\documentclass{PoS}
                                                                                               
\def\beq{\begin{equation}}
\def\eeq#1{\label{#1}\end{equation}}
\def\eeqn{\end{equation}}
\def\beqa{\begin{eqnarray}}
\def\eeqa#1{\label{#1}\end{eqnarray}}
\def\eeqan{\end{eqnarray}}

\def\leqn#1{(\ref{#1})}

\title{Dark Matter Identification using Gamma Rays from Dwarf Galaxies }

\ShortTitle{Dark Matter Identification using Gamma Rays from Dwarf Galaxies }

\author{\speaker{Bibhushan SHAKYA}%
	\\
        Laboratory of Elementary Particle Physics, Cornell University, Ithaca, New York 14853, USA\\
       E-mail: \email{bs475@cornell.edu}}

\author{Maxim PERELSTEIN\\
         Laboratory of Elementary Particle Physics, Cornell University, Ithaca, New York 14853, USA\\
        E-mail: \email{mp325@cornell.edu}}

\abstract{ If the positron fraction and combined electron-positron flux excesses recently observed by PAMELA, Fermi and HESS have a dark matter origin, final state radiation (FSR) photons from dark matter annihilation into lepton-rich final states may be detected with observations of satellite dwarf galaxies of the Milky Way by ground-based atmospheric Cherenkov telescopes (ACTs). We find that current and near-future ACTs have excellent potential for such detection, although a discovery cannot be guaranteed due to large uncertainties in the distribution of dark matter within the dwarfs. We find that models predicting dark matter annihilation into two-lepton final states and those favoring four-lepton final states (as in, for example, ``axion portal" models) can be reliably distinguished using the FSR photon spectrum once measured, and the dark matter particle mass can also be accurately determined.}

\FullConference{Identification of Dark Matter 2010\\
                July 26 - 30 2010\\
                University of Montpellier 2, Montpellier, France}

\begin{document}

\section{Introduction : Final State Radiation, Dwarf Galaxies, ACTs} 

Recent measurements of the positron fraction in cosmic rays in the 10--80 GeV range by PAMELA~\cite{pamela} and the combined electron-positron flux up to the TeV scale by Fermi~\cite{fermi} and HESS ~\cite{hess} indicate excesses inconsistent with conventional astrophysical background. These excesses can be explained, among other possibilities, by dark matter with an annihilation cross section of $\langle \sigma v\rangle \sim3\times10^{-23}$ cm$^3$s$^{-1}$ in the Milky Way and a mass of 1--3 TeV, provided annihilation is predominantly into electrons or muons~\cite{patrick,bestfits}.

If these excesses indeed have a dark matter origin, accompanying signals are expected in the form of energetic gamma rays. For leptophilic dark matter, the high energy end of the gamma ray signal is dominated by final state radiation (FSR). Dwarf galaxies --- made up almost entirely of dark matter, with no detected neutral or ionized gas, minimal dust, no magnetic fields, and little or no recent star formation activity --- are favorable targets for searches for such gamma rays. And since dark matter gamma ray signals from dwarf galaxies are mainly constrained by low statistics, atmospheric Cherenkov telescopes (ACTs), with typical effective areas $\sim 10^5$ times that of Fermi, offer a distinct advantage. In this work, we focus on the prospects of detecting FSR from dark matter annihilation with ACT observations of dwarf galaxies. We also investigate the important question of whether the ACTs can measure the FSR photon spectrum precisely enough to distinguish between different leptophilic dark matter models. For details, please refer to \cite{fsrpaper}.

\textit{Final state radiation (FSR) ---} FSR is present whenever dark matter annihilates to charged particles, as is the case for the following three leptophilic models motivated by fits to PAMELA, Fermi, and HESS data \cite{patrick,bestfits} (with dark matter denoted by $\chi$):\\
(i) Model A: $\chi\chi\rightarrow\mu^+\mu^-$.~~~~ (ii) Model B: $\chi\chi\rightarrow\phi\phi\rightarrow4e$.~~~~(iii) Model C: $\chi\chi\rightarrow\phi\phi\rightarrow4\mu$.\\
$\phi$ denotes an intermediate ``portal" particle, with mass taken to be of order 1 GeV.  Factorization theorems ensure that the energy spectrum of the FSR photons is, to leading order in $m_l/m_{\chi}$, independent of the details of the annihilation process, allowing for quasi-model-independent predictions. For annihilation into a lepton-antilepton pair $l\bar{l}$ as in model A, the FSR flux within the leading log approximation is \cite{robust}
\beqa
\frac{d\Phi_{FSR}}{dx}=\Phi_0\left(\frac{\left<\sigma v\right>}{1pb}\right)\left(\frac{100~{\rm GeV}}{m_{\chi}}\right)^3\frac{1+(1-x)^2}{x}\,\log\left(\frac{4m_\chi^2(1-x)}{m_l^2}\right)\,J,
\label{2state}\\
J=\frac{1}{8.5~{\rm kpc}}\left(\frac{1}{0.3~{\rm GeV/cm}^3}\right)^2L,~~~~L= \int d\Omega\int_{l.o.s.}\rho^2 dl.
\label{jfactor}
\eeqa{factors}
Here $x=2E_{\gamma}/\sqrt{s}=E_{\gamma}/m_{\chi}$, $\Phi_0=1.4\times 10^{-14}$ cm$^{-2}$s$^{-1}$GeV $^{-1}$, and $J$ is a dimensionless factor that carries all the astrophysics information;\footnote{Eq. \leqn{2state} and similar formulas below are equally applicable for decaying dark matter, with an appropriate redefinition of the $J$ factor. However, we do not consider decaying dark matter in this paper since the resulting FSR signals from dwarf galaxies are generally too weak to be detected.} astrophysical uncertainties, therefore, do not affect the spectrum of final state radiation. The spectrum in Eq. (\ref{2state}) features a characteristic ``edge" at the dark matter mass \cite{robust}.

For 4-body annihilation as in models B and C, the FSR spectrum is \cite{robust}
\beq
\frac{d\Phi_{FSR}}{dx}=\Phi_0\left(\frac{\left<\sigma v\right>}{1pb}\right)\left(\frac{100~{\rm GeV}}{m_{\chi}}\right)^32\frac{2-x+2x\log x-x^2}{x}\,\log\left(\frac{m_\phi^2}{m_l^2}\right)\,J.
\eeq{4state}
For models A and C, annihilation is to muons, and final state radiation form the subsequent decay of the muon should also be taken into account. Relevant formulas for this contribution are as given in \cite{rouven, fsrpaper}. Typically, FSR off muons from the annihilation process remains dominant unless $m_{\phi}\sim m_{\mu} $\cite{rouven}. In this work we fix our parameters to $\left<\sigma v\right>=3\times10^{-23}$ cm$^3$s$^{-1}$, $m_{\chi}=3$ TeV, and $m_{\phi}=1$ GeV, as favored by fits to PAMELA, Fermi, and HESS data. It should be kept in mind that lower velocities in dwarf galaxies can lead to larger cross sections via greater Sommerfeld enhancement. 

\textit{Dwarf galaxies ---} 
We use the following dwarf galaxies, which have been known to be promising candidates for dark matter searches, in our analysis: \\
\indent Draco (18.63~$\pm$~0.60) ~~~~~~~~~ Ursa Minor (18.79~$\pm$~1.76)\\
\indent Willman 1 (19.55~$\pm$~0.98) ~~~~~ Segue 1 (19.0~$\pm$~0.6)\\
The number in parenthesis is $\log_{10}(L\times$\,GeV$^{-2}$cm$^{5}$), where $L$ is the astrophysical factor as defined in Eq. (\ref{jfactor}) and calculated in \cite{rouven, newsegue} \footnote{The astrophysical factor for Segue 1 listed here represents an updated value \cite{newsegue} that was not available at the time of writing of \cite{fsrpaper}; all plots and discussions in the following sections use this updated value.}; it should be noted that the uncertainties on these astrophysical factors are extremely large at present. The Sloan Digital Sky Survey has recently discovered many new dwarf galaxies, and since only a small region of the galactic neighborhood has been completely surveyed, several hundred more low-luminosity, dark matter dominated dwarf galaxies might still be discovered in the future. It would be straightforward to apply the analysis of this paper to any promising new dwarf  that may be discovered, once the distribution of dark matter is mapped out to allow for at least an approximate determination of its $L$ factor.

\textit{Atmospheric Cherenkov telescopes (ACTs) ---}
The key parameters governing the sensitivity of an ACT in observations of dwarf galaxies are its effective area $A_{eff}$, energy resolution $\epsilon$, and energy threshold. There are several currently operational (eg. MAGIC, VERITAS) and near future (eg. CTA) ACTs relevant for indirect dark matter detection; we refer to \cite{fsrpaper} for more details, references, and individual key parameter values. Typically, current ACTs have $A_{eff}\sim10^9cm^2$ and $\epsilon\sim 0.15$, while future instruments are expected to reach $A_{eff}\sim10^{10}cm^2$ and $\epsilon\sim 0.10$. The typical instrumental energy threshold for ACT's is about 200 GeV.

\begin{figure}[t]
\centering
\includegraphics[width=3.5in,height=1.8in]{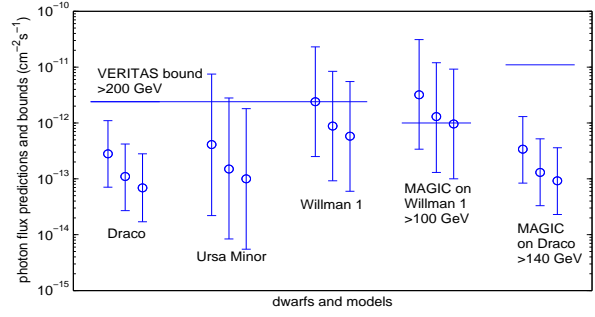}
\caption{Comparison of experimental bounds with predictions from theory. The horizontal lines represent the bounds from experimental observations (\cite{fsrpaper} and references therein). The three vertical bars for each search are the corresponding predictions of models A (left bar), B (center), and C (right), using the astrophysical factors as listed in the text, with circles denoting central values.}
\label{bounds}
\end{figure}

\textit{Previous observations and upper bounds ---}
The dwarf galaxies mentioned above have been observed by ACTs without any positive detection, resulting in upper bounds on high energy gamma ray flux from dark matter annihilation or decay in these galaxies. Figure \ref{bounds}, which compares these experimental bounds with theoretical predictions from the three leptophilic models of interest in this paper, shows that the predictions are consistent with the observed null results within the uncertainties in the astrophysical factors; for more details on bounds from individual observations, the reader is referred to \cite{fsrpaper} and references therein.  

\textit{Backgrounds ---}
Since the dwarf galaxies themselves are not expected to contain significant sources of hard gamma rays of astrophysical origin, the background can be effectively measured by looking at a region of the sky close to the dwarf (called the OFF region); subtracting the OFF region flux from the flux in the ON region (which contains the dwarf) eliminates the background up to statistical fluctuations. Astrophysical backgrounds can be estimated with standard extrapolations of charged lepton, hadron, and gamma ray spectra \cite{bergstrom, robust}; the hadronic background depends on the hadron rejection capabilities of the instrument ( see \cite{fsrpaper} for a more detailed treatment). Gamma rays from dark matter annihilation within the Milky Way also contribute to this background. In addition to FSR, these come from inverse Compton scattering (ICS) of starlight and CMB photons off energetic leptons from dark matter annihilation; we estimate the ICS contribution using the semi-analytic formalism in \cite{positrons}.

\section{Detection Prospects}
Standard requirements for detection are (i) the excess in the ON region relative to the OFF region exceeds 3$\sigma$ (or 5$\sigma$), and (ii) more than 25 (signal) events (after background subtraction):
\begin{eqnarray}
{\mathrm{Significance}}\,=\,\frac{\Phi_{\gamma}A_{\rm eff}t}{(A_{\rm eff}t)^{1/2}(d\Phi_{\rm bg}/d\Omega\times\Delta\Omega)^{1/2}}\geq 3 \mathrm{~(or ~5)},\label{condition1}\\ \nonumber \\
{\mathrm{Number ~of ~signal~events}}\,=\,\Phi_{\gamma}A_{\rm eff}\,t \geq 25,
\label{condition2}
\end{eqnarray}
Here $t $ is the observation time, $\Phi$ denotes gamma ray flux, and $\Omega$ denotes solid angle. For our estimates we ignore systematic errors and assume background subtraction with ON and OFF regions of the same size. To improve sensitivity, it is useful to choose an energy threshold that maximizes the ratio $\Phi_{\rm signal}/\sqrt{\Phi_{\rm bg}}$. The optimum energy threshold is found to lie between 200 and 700 GeV depending on the model \cite{fsrpaper}; for our estimates we use two common values, 200 GeV and 500 GeV, for all three models to allow direct comparisons between models.

\begin{figure}[t]
\centering
\begin{tabular}{cc}
\includegraphics[width=3in,height=1.8in]{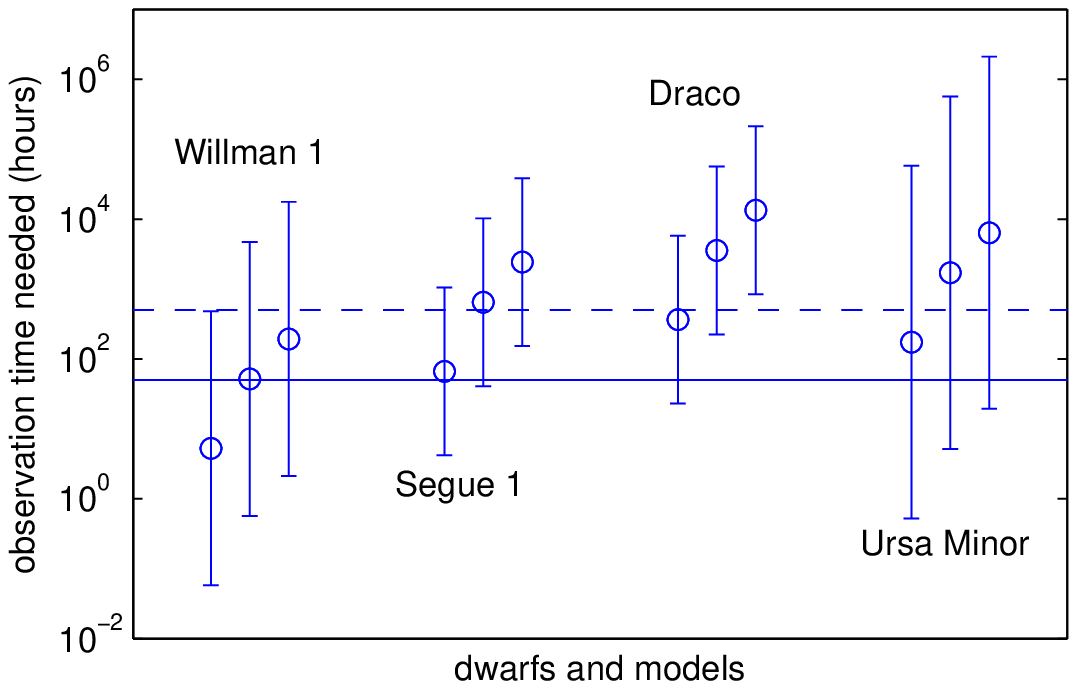}&
\includegraphics[width=3in,height=1.8in]{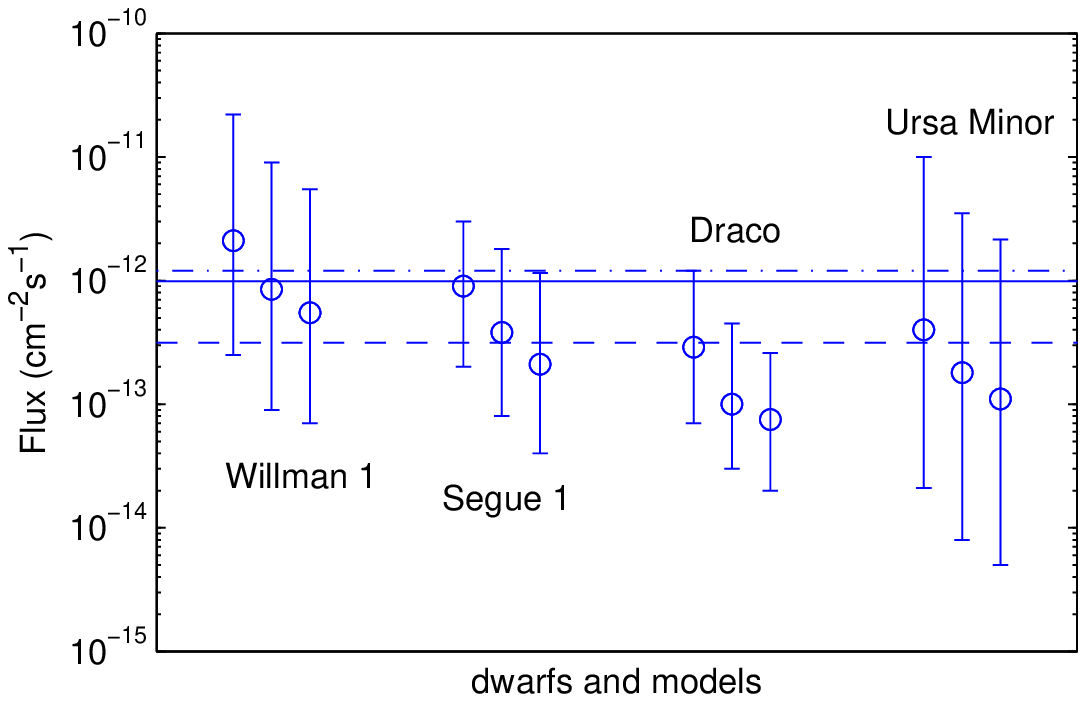}\\
\end{tabular}
\caption{Left: Observation times needed for 3$\sigma$ detection with MAGIC parameters. 
The three points for each dwarf correspond to (from left to right) models A, B and C with a 200 GeV energy threshold. Vertical bars correspond to the uncertainties in the astrophysical factors of the dwarfs, with the circles corresponding to central values. The solid horizontal line denotes 50 hours of observation time with parameters of the MAGIC telescope. The dashed line, at 500 hours, is equivalent to 50 hours of observation with an order of magnitude increase in the effective area, \textit{i.e.} CTA parameters. Right: Integrated fluxes above 200 GeV; the dot-dashed, solid, and dashed lines correspond to approximate sensitivities of VERITAS, MAGIC, and CTA respectively, for 50 hours of observation time.}
\label{sensitivities}
\end{figure}

Figure~\ref{sensitivities} (left) shows the length of observation time needed for a 3$\sigma$ detection of the FSR signal from each dwarf for each annihilation model, for an energy threshold of 200 GeV. Figure~\ref{sensitivities} (right) shows the minimum integrated flux above 200 GeV that can be detected at the $3\sigma$ level in 50 hours of observation for a few ACTs, and the flux predicted by each model for various dwarfs. It is clear from these plots that the uncertainties in the astrophysical factors, and therefore observation times and sensitivities required for a positive signal, span several orders of magnitude; as a result, no model is ruled out, and no dwarf is guaranteed to give an observable signal.  These estimates only allow us to conclude that, for current and future ACTs, detection of FSR from dark matter annihilation from the above dwarfs is \textit{likely}, but not \textit{guaranteed}. Further astronomical observations of the dwarfs should reduce the uncertainty in the $L$ factors, allowing more precise predictions to be made.

\section{Model Identification : Case Studies}

We investigated the prospects of identifying the dark matter model and its parameters based on an observed FSR gamma ray signal, and the effects of changes in the energy threshold, the dark matter mass, the hadron rejection capabilities of the instrument, and the strength of the signal. 

Our approach for these studies is as follows. Assuming one of the models (A, B, or C) is realized in nature, we generated a set of random ``data points" distributed in energy in accord with the theoretical predictions of this model, incorporating the energy resolution of the instrument. The total number of data points corresponds to the prediction of the model for a particular source and telescope parameters. These data points were then binned, choosing the bin width to be approximately double the energy resolution of the ACT being considered. The dark matter mass will in practice be unknown, but the final bin can always be made large enough that $E_\gamma=m_\chi$ can be assumed to fall in this bin.

For each data set, two background samples, corresponding to the ON and OFF regions, were generated. The number of background events in each bin was calculated as the difference between the event counts in the ON and OFF samples in that bin. A fit to the binned data points was performed with all three models, varying the parameters in each model to minimize $\chi^2/d.o.f.$. The three optimal $\chi^2/d.o.f.$ values obtained in this way were compared to each other, and the model with the smallest value was declared the ``best fit" to the data. We performed this procedure for 100 randomly generated data sets for each of the models A, B, and C. The analysis was performed for two sets of ACT parameter values: $A_{\rm eff}=10^9$ cm$^2$ and 15\% energy resolution, representative of the reach of current instruments, and $A_{\rm eff}=10^{10}$ cm$^2$ and an energy resolution of 10\%, representative of the reach of future instruments. 

\textit{3$\sigma$ or 5$\sigma$ detection} --- As an example, we provide the results for the case when the FSR flux is on the threshold of detection. We performed this analysis assuming the weakest possible signals that can be detected at 3$\sigma$ or 5$\sigma$ significance above 500 GeV. Table~\ref{tab:case5} shows the fit results for current telescope parameters; results for future telescope parameters are similar.
\begin{table}[h!]
\begin{center}
\begin{tabular}{|c|c|c|c|c|}\hline
``True" model & model A,B,C & Best fit WIMP  & Best fit  &$\chi^2/d.o.f.$\\
& as best fit & mass (GeV) &cross-section (pb)&\\ \hline
\underline{3$\sigma$ detection}&&&&\\
model A & 71, 12, 17 & $3092\pm531$&$1235\pm265$& $~1.17\pm0.71$~ \\
model B & 2, 62, 36 & $3041\pm524$ & $1236\pm275$ &~ $1.74\pm1.62$~ \\
model C & 0, 49, 51 & $3051\pm503$ & $1238\pm650$& $~1.26\pm0.77$~\\
\hline
\underline{5$\sigma$ detection}&&&&\\
model A & 83, 9, 8 & $3146\pm433$&$1087\pm182$& $~1.32\pm0.76$~ \\
model B & 2, 64, 34 & $3117\pm471$ & $1148\pm238$ &~ $1.51\pm0.97$~ \\
model C & 3, 17, 80 & $3065\pm489$ & $1005\pm629$& $~1.44\pm0.95$~\\
\hline
\end{tabular}
\end{center}
\caption{Fit results for the weakest possible signal detectable at 3$\sigma$ and 5$\sigma$ levels by the current ACTs.}
\label{tab:case5}
\end{table}

The first entry, the distribution of ``best fit" models, contains three numbers, corresponding to the number of data sets for which models A, B, and C respectively gave the best fit. For the other parameters we also list the statistical error bar reflecting the variation of the best-fit values and the $\chi^2/d.o.f.$ of the best fit among the 100 data sets. The fit results show that the overall success rate for correctly identifying the model in these scenarios is 61\% and 75\% for 3$\sigma$ and 5$\sigma$ detection respectively. The success rates of correctly identifying the signal as a 2-body or 4-body final state are excellent: 90\% and 93\% respectively. Best fit values obtained for the WIMP mass and annihilation cross section are fairly accurate.

The reader is referred to \cite{fsrpaper} for detailed results for all cases studied. Lowering the energy threshold from 500 GeV to 200 GeV did not seem to have any clear benefits. For a dark matter of a lower mass (1 TeV instead of 3 TeV), there was some improvement in distinguishing between 2 and 4 body annihilation channels for current telescope parameters. Improving the hadron rejection capabilities of telescopes also appeared to improve model identification efficiency (from $\sim$ 70\% to $\sim$ 82\%). Overall, the success rate for proper model identification was found to be fairly robust with respect to changes in energy threshold, WIMP mass, energy resolution, and hadron rejection capabilities of the telescope, and appeared to depend mainly on the signal to background ratio. 

All cases considered in our analysis shared the following common features. (1) The 2$\mu$ channel was very clearly identified because of its `edge' feature. (2) The two 4-body final state channels were not easily distinguishable from each other because their spectra are very similar. (3) Best fit values for the dark matter mass and annihilation cross section were in excellent agreement with their ``true" values. The mass was better reconstructed with future telescope parameters because of superior energy resolution. The accuracy of the mass and cross section determination was particularly impressive (of order 1\%) in the 2$\mu$ channel, again presumably due to the sharp edge feature at $E_\gamma=m_\chi$. (4) Future telescope parameters showed a clear improvement over current parameters in terms of correct identification of the model. This was mainly due to the increase in effective area, which results in a greater number of events and a better significance.

\section{Concluding remarks}
If the PAMELA, Fermi and HESS anomalies have their origin in leptophilic dark matter annihilation, current and near-future ACTs have an excellent chance of observing the accompanying final state radiation from dwarf galaxies. Unfortunately, lack of precise knowledge of the distribution of dark matter in the dwarfs makes the signal flux predictions highly uncertain. If a signal is observed, the measured gamma ray spectrum can likely be used to identify the correct annihilation channel and dark matter mass --- a general conclusion that holds for a range of signal strengths, dark matter masses, energy thresholds, and instrument parameters --- paving the way to a better understanding of the microscopic nature of dark matter.

\vskip0.5cm
\noindent{\large \bf Acknowledgments} 
\vskip0.3cm

This research is supported by the U.S. National Science Foundation through grant PHY-0757868 and CAREER award PHY-0844667.

\end{document}